\documentclass[pra,amsmath,amssymb,superscriptaddress,showpacs,twocolumn]{revtex4-1}
\usepackage{graphicx}
\usepackage{amsmath}
\usepackage{bm}
\usepackage{graphicx}
\usepackage{dcolumn}
\usepackage{bm}
\usepackage{colordvi}
\usepackage{color}
\usepackage{ulem}
\newcommand{\beq}{\begin{equation}}
\newcommand{\eeq}{\end{equation}}
\newcommand{\beqa}{\begin{eqnarray}}
\newcommand{\eeqa}{\end{eqnarray}}

\begin{document}

\title{Spin dynamics in tunneling decay of a metastable state}

\author{Yue Ban}
\affiliation{Department of Physical Chemistry, The University of the Basque
Country UPV/EHU, 48080 Bilbao, Spain}
\author{E. Ya. Sherman}
\affiliation{Department of Physical Chemistry, The University of the Basque
Country UPV/EHU, 48080 Bilbao, Spain}
\affiliation{IKERBASQUE Basque Foundation for
Science, Bilbao, 48011 Bizkaia, Spain}

\date{\today}

\begin{abstract}
We analyze spin dynamics in the tunneling decay of a metastable localized
state in the presence of spin-orbit coupling. We find that the spin
polarization at short time scales is affected by the initial state while at
long time scales both the probability- and the spin density exhibit
diffraction-in-time phenomenon. We find that in addition to the tunneling time
the tunneling in general can be characterized by a new parameter, the tunneling length.
Although the tunneling length is independent on the spin-orbit coupling, it can be accessed by the spin rotation
measurement.
\end{abstract}


\maketitle

\section{Introduction}

Spin-orbit (SO) coupling, interaction of particle spin or pseudospin
with the orbital motion, provides an
efficient way to control and manipulate spin, charge, and mass transport.
Two different kinds of SO coupled systems attract a great deal of interest
due to the known and yet unexplored variety of phenomena they can demonstrate and due to
their possible applications in quantum technologies.
One class of systems, investigated for more than five decades by now,
is semiconductors and semiconductor-based
nanostructures \cite{review_spintronics,Amasha,Stano,Stano2,Romo,Malajovich,Awschalom}.
The other class, lavished attention only recently, is cold atoms and Bose-Einstein
condensates, where by engineering external
optical fields one can couple orbital motion to the pseudospin degree of
freedom \cite{Stanescu,Liu,Wang,Lin,Campbell} and cause Dresselhaus and Rashba types
of SO coupling similar to that in solids.

In quantum systems of interest the tunneling plays an important role and either completely
determines or strongly influences the particle dynamics. At certain conditions the tunneling
rate depends on the spin of the particle \cite{Amasha,Stano,Glazov05,Sherman}.
The understanding of the tunneling is the key for the understanding
of the transient processes in a broad variety of systems, including, e.g.
charge transport in molecular nanostructures \cite{transient}.
One of the key issues in the tunneling theory is the evaluation of the time spent
by the particle in the classically forbidden regions.
Similar to the scattering problem of propagating wave packets, the question of the tunneling time for the decay of a metastable system \cite{decay1,decay2,decay3,decay4,decay5} may also be posed. The spin-dependent effects based on the Larmor clock
concept for potential barriers \cite{Baz,Rybachenko}, Hartman effect in graphene
\cite{Beenakker}, and effective exchange fields in semiconductors \cite{Appelbaum2}
can provide a measure of this time.
In classically forbidden regions there is not only
precession but also a rotation of the magnetic moment into
the magnetic field direction and the time of the interaction with the barrier
is closely related to this rotation \cite{Buttiker,Buttiker-Landauer}.
The effects of SO coupling on the tunneling through semiconductor
quantum-well structures with a lateral potential barrier \cite{SOC-tunneling,SOC-tunneling2} provide another tool to
utilize electron spin modifying the charge transport in nonmagnetic systems.

Here we investigate the spin-dependent tunneling of a state initially localized
in the potential at short  and long compared to the state lifetime time scales \cite{tunneling1}.
This paper is organized as follows. In Sec. II we introduce the model Hamiltonian and formulate
the physical observables of interest. We use the SU(2) spin rotation to gauge
out spin-orbit coupling and restore its effects in the calculation of the observables
by inverse transformation.
In Sec. III we study the dynamics at short time scale, concentrating on spin oscillations,
decay, and escape from the localizing potential.
In Sec. IV we study long-term dynamics in the far-field zone.
The diffraction in time in spin density is observed at a long distance, where
SO coupling forms a precursor in the propagating density. We show that
the tunneling can be characterized not only by time, but also by a certain
length parameter, which can be accessed by detecting the spin precession
due to the SO coupling. Conclusions summarize the results.

\section{Model for spin-dependent tunneling}

\begin{figure}[ht]
\begin{center}
\scalebox{0.5}[0.55]{\includegraphics{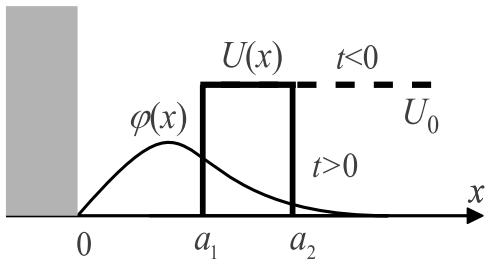}}
\caption{Time-dependent potential $U(x)$: $U_1(x)$ at $t<0$ and $U_2(x)$ when $t>0$.
$U_1(x)$ is a step in the positive halfplane, that is zero from $x=0$ to
$x=a_1$ and $U_0$ from $x=a_1$ to infinity, while $U_2(x)$ is a barrier of the height $U_{0}$ extended from $x=a_1$ to $x=a_2$.
The potential is always infinite in the negative halfplane. The barrier width is defined as $d\equiv a_{2}-a_{1}$.}
\label{model}
\end{center}
\end{figure}

As a model we consider shown in Fig.\ref{model} time-dependent potential $U(x,t),$ infinite at $x< 0$,
with the time-dependence:
\begin{equation}
U\left(x,t\right)=U_1(x)  ~~ (t<0){,}\quad
U\left(x,t\right)=U_2(x) ~~ (t>0){.}
\end{equation}
At $t\leq 0$ the potential holds bound states
with wave functions $\varphi_j(x)$ and energy $E_j$;
it changes at $t>0$ from a step to a barrier to allow the tunneling.
Such time-dependent potential can be produced by a recently developed
technique \cite{potential} where a moving laser beam ``paints'' a broad
variety of coordinate- and time-dependent potentials.

With SO coupling taken into account, the total one-dimensional Hamiltonian is
\begin{eqnarray}
\label{hamiltonian1}  \hat{H}
=\frac{1}{2m} \left(\hbar\hat{k}_x + \hat{A}_x \right)^2+U(x) - \frac{p_{\rm so}^2}{2m},
\end{eqnarray}
with $\hat{k}_x=-i \partial/\partial x$, the vector potential
$\hat{A}_{x} =p_{\rm so}\hat{\sigma}_x$, $p_{\rm so}\equiv m\alpha/\hbar$,
where $\alpha$ is the Dresselhaus SO coupling constant, and $\hat{\sigma}_x$ is the corresponding
Pauli matrix. The spatial scale of spin precession is characterized by the length $2\xi=\hbar/p_{\rm so}$.
The typical values of $\xi$ for different systems with SO coupling can be of interest.
For (i) electrons in semiconductor GaAs nanostructures \cite{review_spintronics}
with $m \approx 6\times 10^{-29}$ g, $\alpha/\hbar \sim 0.5\times10^{6}$ cm/s,
(ii) for $^{6}{\rm Li}$ atoms \cite{Liu} with $m \approx 10^{-23}$ g, $\alpha/\hbar\sim 10$ cm/s, and
(iii) for $^{87}{\rm Rb}$ atoms \cite{Campbell} with $m \approx 1.5\times10^{-22}$ g, $\alpha/\hbar\sim 0.3$ cm/s,
we obtain $\xi\sim 10^{-5}$ cm. It is interesting to mention that although masses and coupling constants
for electrons in nanostructures and cold atoms are very different,
the resulting precession length is on the same order of magnitude for all
these systems. Since typical localization scale of electrons in nanostructures $a_{1}$ is on the order
of $10^{-6}$ cm, they are in the weak-coupling regime with $\xi/a_{1}\gg1$. Cold atoms,
however, are localized on the scale of the order visible light wavelength, that is of $10^{-4}$ cm, and
the strong SO coupling regime with $\xi<a_{1}$ can be achieved there. The characteristic
timescale corresponding to the particle motion inside the potential, $ma_{1}^{2}/\hbar$,
is on the order of 0.1-1 ps for electrons in nanostructures and 0.1-1 ms for cold atoms.

From now on we use the system of units with $\hbar\equiv 1$ and particle effective mass $m\equiv 1$.
The wave function corresponding to $\hat{H}$ is $\bm{\psi}(x,t)$ with the initial state set as
\begin{eqnarray}
\label{wave function0}  \bm{\psi}(x,0) =  \chi \varphi_j(x),
\end{eqnarray}
and the spinor $\chi=\left[1,0\right]^{T}$ corresponds to the  $z$-axis
orientation. Such a function can be produced by applying a magnetic field along the
$z$-axis for an electron or by a special design of optical field in  the case
of a neutral bosonic atom \cite{Campbell}.
At $t>0$, neither the orbital wave function $\varphi_j$ is the eigenstate, nor
$\chi$ is the eigenspinor of Hamiltonian $\hat{H}$.
The initial state begins to evolve at $t>0$ with spin precession
and the probability to find the particle inside the potential decreases simultaneously.

We use a gauge transformation $\hat{\tilde{H}}=\bm{S}\hat{H}\bm{S}^{-1}$
with a unitary matrix  $\bm{S}=\exp(i \hat{\sigma}_{x}x/2\xi)$ to gauge away the SO coupling.
The gauge transformation shifts $\hat{k}_x$ by $\hat{\sigma}_x/2\xi$ and turns SO coupling
into a constant, with the wave function
evolving in time transformed as:
\begin{eqnarray}
\label{wave function change}
\bm{ \tilde{\bm{\psi}}}(x,t) = \bm{S} \bm{\psi}(x,t)=\left[
\begin{array}{c} \tilde{{\psi}}_{1}(x,t)\\
                 \tilde{{\psi}}_{-1}(x,t)
\end{array}\right],
\end{eqnarray}
with the upper and lower components
\begin{eqnarray}
\label{wave function new2}
\tilde{\bm{\psi}}_{\sigma}(x,t)=
\int_{0}^\infty G_{\sigma}(k) \phi_k(x) \exp \left(-\frac{i k^2 t}{2}\right) dk,
\end{eqnarray}
and the coefficients are
\begin{eqnarray}
\label{coefficient}G_{\sigma}(k) =\int_0^\infty \tilde{\bm{\psi}}_{\sigma}(x,0)\phi_k(x) dx.
\end{eqnarray}
Here, $\phi_k(x)$ is the eigenstate of the Hamiltonian corresponding to $U_2(x)$:
\begin{eqnarray}
\label{eigenstateinbarrier}\ \phi_k(x)= \left\{
\begin{array}{ll}
C(k) \sin(k x),  &(0<x<a_1) \\
\\
D(k) e^{-\kappa_{k}x} + F(k) e^{\kappa_{k}x}, &(a_1<x<a_2) \\
\\
\sqrt{{2}/{\pi}} \sin[k x+\theta(k)], &(x>a_2)
\end{array}
\right.
\end{eqnarray}
normalized as $\langle\phi_{k'}|\phi_{k}\rangle=\delta(k-k')$.
The coefficients $C(k)$, $D(k)$, $F(k)$ and the phase $\theta(k)$ satisfy the
boundary conditions of the potential $U_2(x)$, and $\kappa_{k}=\sqrt{2U_0-k^2}$.
In the tunneling regime $k<\sqrt{2U_0}$, while in the propagating regime $k>\sqrt{2 U_0}$, and $i\kappa_{k}$
is substituted by $q=\sqrt{k^2-2U_0}$.

The initial wave function corresponding to Hamiltonian $\hat{\tilde{H}}$ can be expressed as
\begin{equation}
\label{initial wave function}  \tilde{\bm{\psi}}(x,0) = \varphi_j(x)\left[\begin{array}{c} \cos(x/2\xi) \\ i\sin(x/2\xi)
\end{array}\right].
\end{equation}
The coefficients $G_{\sigma}(k)$ become
\begin{eqnarray}\label{coefficient1}
G_{1}(k) = \int_0^\infty\cos(x/2\xi) \varphi_j(x)\phi_k(x) dx,\\
G_{-1}(k) = i\int_0^\infty\sin(x/2\xi)\varphi_j(x)\phi_k(x) dx.
\label{coefficient2}
\end{eqnarray}

In what follows, we investigate the  particle motion by calculating the physical
observables such as probability density, spin density, and
spin polarization defined as:
\begin{eqnarray}\label{density}
&&\rho(x,t) = \bm{\psi}^\dag (x,t) \bm{\psi}(x,t),\\
&&\sigma_{i}(x,t) = \bm{\psi}(x,t)^\dagger \hat{\sigma}_i \bm{\psi}(x,t),\\
&&p_{i}(x,t)=\frac{\hat{\sigma}_{i}(x,t)}{\rho(x,t)},
\end{eqnarray}
respectively. As the spin rotates around the effective SO coupling field along the
$x$-direction, the integral of $\sigma_{x}(x,t)$-component over the $x>0$ half axis
is conserved. For this reason, we investigate spin density and spin polarization in the more
informative $y-$component.

\section{Spin dynamics at short times}

To address the short-term dynamics on the time less than the lifetime
of the initial bound state \cite{tunneling1}, we study two relevant quantities.
First quantity is $\sigma_y(a_2,t)$, the spin density at the exit
of the barrier $a_2$.  The other one, $p^{[w]}_y(t)$, defined as
\begin{eqnarray}
\label{spin polarization in y}
p^{[w]}_y(t) =
\frac{\displaystyle{\int_0^{a_1}}\bm{\psi}^{\dagger}(x,t)\sigma_y\bm{\psi}(x,t)dx}
{\displaystyle{\int_{0}^{a_1}}\bm{\psi}^{\dagger}(x,t)\bm{\psi}(x,t)dx},
\end{eqnarray}
being the spin polarization $y$ component in the potential, represents the
integrated spin dynamics inside the potential. The denominator
of Eq.(\ref{spin polarization in y}) is the probability to find the particle
inside the potential, which decays to $1/e$ of its initial value
at the lifetime of the metastable state.

\begin{figure}[ht]
\scalebox{0.65}[0.65]{
\includegraphics*{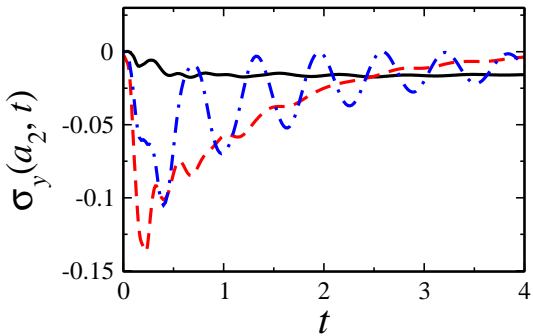}}
\caption{(Color online) Comparison of time evolution of $\sigma_y(a_2,t) $ at the right edge of the barrier $a_2$
with different initial states,
the ground state $\varphi_0$ (solid), the first excited state $\varphi_1$ (dashed) and the linear
combination $\varphi_{\textrm{com}}=(\varphi_0+\varphi_1)/\sqrt{2}$ (dot-dashed),
provided by the barrier $d=0.4$, $U_0=16$, and $\xi=0.5$.
Negative $\sigma_y(a_2,t)$ is related to the direction of spin precession determined by the sign
of coupling constant $\alpha$.}
\label{Sya2}
\end{figure}

We consider the evolution of three initial orbital states, as shown in Fig.\ref{Sya2}
for the spin component at the edge.
We choose $U_0=16$ so that there are two bound states, the ground state $\varphi_0$ and the first excited
state $\varphi_1$, and use dimensionless $a_1\equiv1$. Here, the initial state has
a strong impact on the time-dependent $\sigma_y(a_2,t)$. It is demonstrated in Fig.\ref{Sya2}
that it takes very short time for $\sigma_y(a_2,t) $ to develop into the minimum from zero, irrespective
of the initial state.  This behavior is similar to the fast development of a plateau in
the outgoing flux after the potential change \cite{tunneling1}.
However, spin density with $\varphi_1$  decays faster than that
with $\varphi_0$. The linear combination of these two bound states
$\varphi_{\textrm{com}}=(\varphi_0+\varphi_1)/\sqrt{2}$ shows
strong oscillations due to the interference between them. As both spin density and probability density in the potential
decay with time, spin polarization tends to a constant at large time.

\begin{figure}[ht]
\scalebox{0.65}[0.65]{\includegraphics*{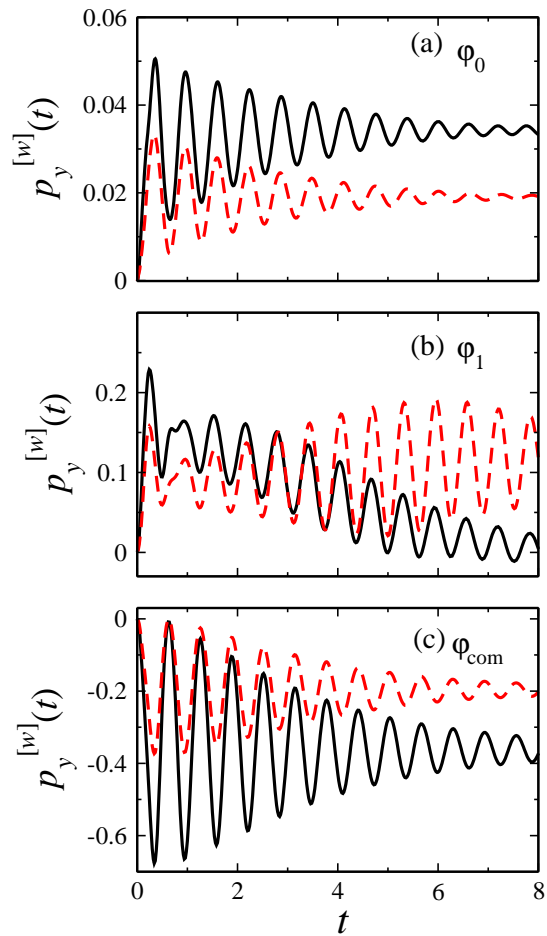}}
\caption{(Color online) Time evolution of $p^{[w]}_y(t)$ in the potential from $0$ to $a_1$, caused by
SO coupling with $\xi=0.5$ (solid) and $\xi = 1$ (dashed), with different initial coordinate states:
(a) $\varphi_0$, (b) $\varphi_1$, and (c) $\varphi_{\textrm{com}}=(\varphi_0+\varphi_1)/\sqrt{2}$,
provided by the same barrier as in Fig.\ref{Sya2}. }
\label{wellPy}
\end{figure}

The tunneling resulted from various initial states has a strong effect on spin polarization,
especially in short-time scales, as shown in Fig.\ref{wellPy}. Spin polarization inside the potential oscillates
between boundaries. The oscillation rate is fast and determined by the
energy difference of the initially bound states. With the time, the contribution of the excited
state rapidly decays due to the fast tunneling, and the spin remains
in the state achieved by the time of the decay of the upper bound state.
The amplitude of resonances for $\varphi_1$ is larger than $\varphi_0$, as the
former possesses larger momentum. As for the linear combination of these two bound states, the spin polarization
is greatly enhanced because of interferences. On the other hand, strength of SO coupling strongly influences the
spin polarization in the potential as illustrated by making comparisons for $\xi=0.5$ and $\xi=1$
parameters.

\begin{figure}[ht]
\scalebox{0.7}[0.7]{\includegraphics*{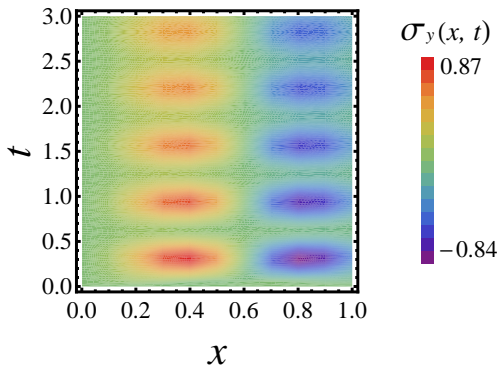}}
\caption{(Color online) Spin density ($y$ component) in the potential vs
time and position for $\xi=0.5$ and $\bm{\psi}(x,t=0)=\chi \varphi_0$.
The parameters of the barrier are the same as in Fig.\ref{Sya2}.}
\label{wellsy}
\end{figure}

Having illustrated the evolution of spin density at the boundary, $\sigma_y(a_2,t)$, and
averaged spin polarization, we can address the details of the spin distribution inside the potential.
As shown in Fig.\ref{wellsy}, this distribution  oscillates
due to the interferences between the ground state and other eigenstates.
The amplitude of oscillations becomes weaker with time as the state decays.

\section{Spin dynamics in the far-field zone}

\begin{figure}[ht]
\scalebox{0.65}[0.65]{\includegraphics*{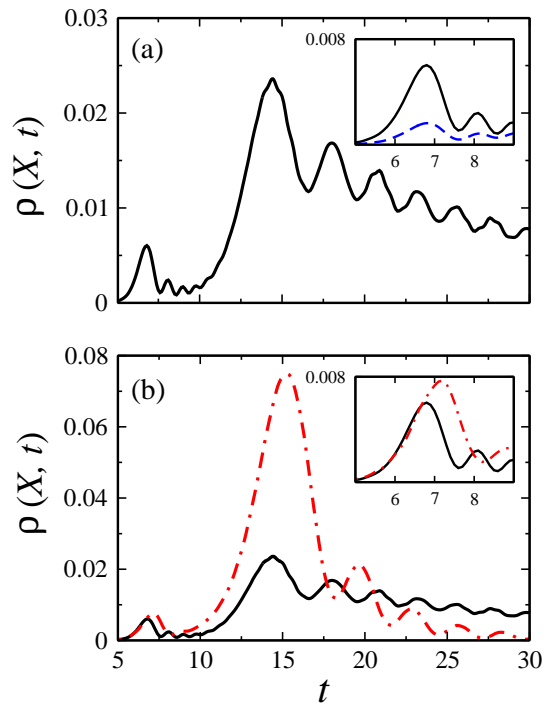}}
\caption{(Color online) (a) Time-dependent density at the observation point
$X=10\pi$ for system parameter $U_{0}=16, d=0.4$, and $\xi=0.5$.
Inset shows the precursor dynamics on a shorter time scale. Dashed line corresponds
to $U_{0}=16, d=0.4$, and $\xi=1$.
(b) Solid line corresponds to $U_{0}=16, d=0.4$, and $\xi=0.5$,
dot-dashed line corresponds to $U_{0}=8, d=0.4$, and $\xi=0.5$.
Inset shows the precursor dynamics on a shorter time scale.}
\label{density:fig}
\end{figure}

\begin{figure}[ht]
\scalebox{0.65}[0.65]{\includegraphics*{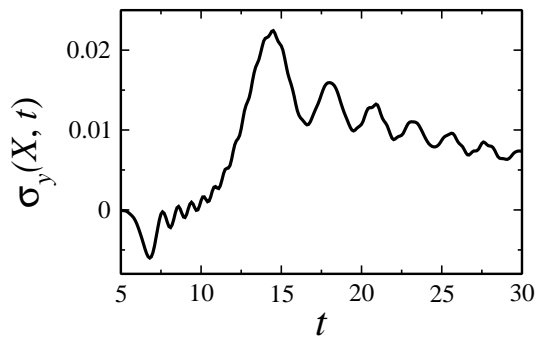}}
\caption{ Spin density in the observation point $X=10\pi$
for $U_{0}=16, d=0.4$, $\bm{\psi}(x,t=0)=\chi \varphi_0$ and $\xi=0.5$. The precursor is made by the contribution of the opposite spin.
Other parameters are the same as in  Fig.\ref{Sya2}.}
\label{sy}
\end{figure}

Here we investigate the long-term spin dynamics,
by considering the probability density $\rho$ and $y$ component of spin density
$\sigma_y(X,t)$ which can be detected at a given long distance $X \gg 1$. The density evolution
for  $\rho(X,t)$ at $X = 10 \pi$ is shown in Fig.\ref{density:fig}
for different SO coupling strength. Under different $\xi$, the time evolutions of probability density show
a strong sharp peak followed by oscillations due to the diffraction-in-time process \cite{Moshinsky,Steane,Campo}.
However, unlike the case without SO coupling \cite{tunneling1,DIT},
for strong couplings (small $\xi$), there exist some oscillations before the sharp peak, where the interference of
two velocities of spin up and spin down components are remarkable. A very interesting feature is the precursor of the
main peak corresponding to the opposite spin. The precursor becomes stronger with the increase in the SO
coupling. The time dependence of $\sigma_y(X,t)$ at $X=10 \pi$ is shown
in Fig.\ref{sy}. Different values of $\xi$ result
in different spin rotation angles and corresponding spin density, provided that $\sigma_y(x,t)$ is detected at the same position.


Spin precession in the $(y,z)$-plane is described by the classical precession angle, $\beta^{\rm [so]}_0 = \Omega t$,
where the precession frequency is $\Omega \approx 2 \alpha k_0 = k_0/\xi$ and the wave packet is detected at the
position $X$ at the time instant $X/v$,
the velocity $v=k_0$. Therefore, the rotation angle is $\beta^{\rm [so]}_0 ={X}/{\xi}$,
independent on time $t$, provided that the wave packet is considered classically.
Ideally, for a free particle propagating from the origin  $x(t=0)=0$, spin polarizations components
are $p_x=0$, $p_y=-\sin \beta^{\rm [so]}_0$, $p_z= \cos \beta^{\rm [so]}_0$. As a result, at $X = n\pi\xi$, the classical spin
polarization should be $p_y=0.$  However, in the tunneling problem we consider, the value of $p_y$
is not zero even at $X = n\pi\xi$.

\begin{figure}[ht]
\scalebox{0.65}[0.65]{\includegraphics*{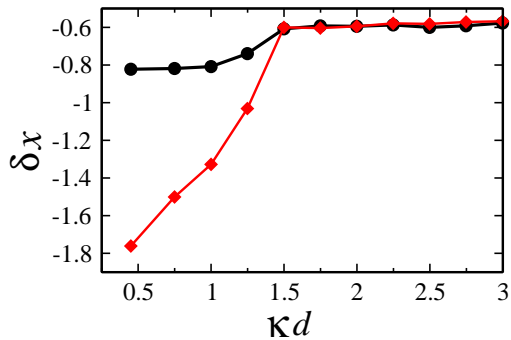}}
\caption{(Color online) Parameter of additional displacement $\delta x$ as the function of the barrier transparency.
Filled circles correspond to $U_{0}=16$ and variable width $d$, the squares correspond to
$d=0.4$ with variable $U_{0}$, and $\xi=10$.}
\label{deltax:tun}
\end{figure}

For the weak SO coupling $\xi\gg1$, where
$\left|{\psi}_{-1}/{\psi}_{1}\right|\ll 1$, the spin polarization can be calculated as:
\begin{eqnarray}
\label{Py correction approximation}
p_y \approx -\sin\frac{x}{\xi} +
2\cos\frac{x}{\xi}{\Im}\frac{{\psi}_{-1}}{{\psi}_{1}}.
\end{eqnarray}
Second term in Eq.(\ref{Py correction approximation}) can be viewed
as a result of a correction to the particle displacement in the form
$p_y =-\sin \left({x}/{\xi}+{\delta x}/{\xi}\right)$,
where, according to Eqs.(\ref{coefficient1}),(\ref{coefficient2}):
\begin{eqnarray}
\label{delta:x}
\delta x = -2\lim_{\xi\rightarrow \infty}\xi\Im
\frac{\displaystyle{\int_{0}^\infty} G_{-1}(k) \exp(-{i k^2 t}/{2}) \phi_k(x) dk}
{\displaystyle{\int_{0}^\infty}G_{1}(k) \exp(-{i k^2 t}/{2}) \phi_k(x) dk}.
\end{eqnarray}
This deviation of $p_y$ from zero at $X=n \pi \xi$ demonstrates that the real precession angle is $\beta^{\rm [so]}=\beta^{\rm [so]}_0+\delta \beta^{\rm [so]}$,
with the correction $\delta \beta^{\rm [so]}$ can be viewed as a result of additional displacement of the particle
$\delta x$ with ${\delta x}={\xi}\delta\beta^{\rm [so]}.$
From Eq.(\ref{delta:x}), we can see that  $\delta x$ is independent of the SO coupling, while
the parameters of the barrier play an important role, as the eigenfunctions $\phi_{k}(x)$ strongly depend on them.
Numerical results also show that $\delta x$ is independent of coordinate and time
if measured at distance $X\gg a_{2}$ at times $t > X/v$.
The tunneling length $\delta x$ is presented in Fig.\ref{deltax:tun}
for two different barriers. This figure shows a clear crossover
from the classical (transparent barrier) to the tunneling (opaque barrier)
regimes, where $\delta x$ shows a saturation, and the additional displacement is universal.
The saturation of the tunneling length for opaque barriers seems similar
to the Hartman effect \cite{hartman} on the group delay in a scattering case \cite{wigner}.
However, after making comparisons of $\delta t\equiv\delta x/v$ and calculated
group delay for a single barrier with parameters presented in Fig.\ref{deltax:tun}, we cannot draw the
unambiguous conclusion on the relevance of these two quantities.

\section{Conclusions}

In conclusion, we investigated the spin dynamics of a tunneling particle initially localized in a potential
in the presence of SO coupling. It is shown that at short-time scales initial states play an important role in
spin polarization while the spin density and probability density possess  diffraction-in-time phenomenon at
long-time scales. We showed that in addition to the time, tunneling can be characterized by
a characteristic length. We use the rotation angle to identify the tunneling length in the presence of weak
coupling. The tunneling length depends on the barrier parameters, being independent on the strength of the spin-orbit coupling.
These effects can be observed in experiments with cold atoms, where the SO coupling is strong
enough to cause spin precession on a relatively short spatial scale.

\section*{Acknowledgement}

This work of EYS was supported by the MCINN of Spain
Grant FIS2009-12773-C02-01, by ``Grupos Consolidados UPV/EHU del Gobierno Vasco" Grant IT-472-10,
and by the UPV/EHU under program UFI 11/55. Y. B. acknowledges financial support
from the Basque Government (Grant No. BFI-2010-255). We are grateful to J.G. Muga and D. Sokolovski
for valuable discussions.

\end{document}